\documentclass[12pt]{article}
\pdfoutput=1
\usepackage[centertags]{amsmath}
\usepackage{amsfonts}
\usepackage{amssymb}
\usepackage{amsthm}
\usepackage{graphicx}
\newcommand{\beq}{\begin{equation}}
\newcommand{\eeq}[1]{\label{#1}\end{equation}}
\newcommand{\bea}{\begin{eqnarray}}
\newcommand{\eea}[1]{\label{#1}\end{eqnarray}}

\begin{document}
\setlength{\topmargin}{-1cm} \setlength{\oddsidemargin}{0cm}
\setlength{\evensidemargin}{0cm}

\begin{titlepage}
\begin{center}
{\Large \bf Holographic Constraints On A Vector Boson}

\vspace{20pt}

{\large Manuela Kulaxizi and Rakibur Rahman}

\vspace{12pt}

Universit\'e Libre de Bruxelles \& International Solvay Institutes\\
ULB-Campus Plaine C.P. 231, B-1050 Bruxelles, Belgium

\end{center}
\vspace{20pt}

\begin{abstract}
We show that holography poses non-trivial restrictions on various couplings of an interacting field.
For a vector boson in the AdS Reissner-Nordstr\"om background, the dual boundary theory is pathological
unless its electromagnetic and gravitational multipole moments are constrained. Among others, a generic
dipole moment afflicts the dual CFT with superluminal modes, whose remedy bounds the gyromagnetic ratio
in a range around the natural value $g=2$. We discuss the CFT implications of our results, and argue that
similar considerations can shed light on how massive higher-spin fields couple to electromagnetism and gravity.
\end{abstract}

\end{titlepage}

\newpage

%%%%%%%%%%%%%%%%%%%%%%%%%%%%%%%%%%%%%%%
\section{Introduction}\label{sec:Intro}
%%%%%%%%%%%%%%%%%%%%%%%%%%%%%%%%%%%%%%%

Any quantum field theory set to describe nature should satisfy a set of physical requirements such as causality,
unitarity etc. Still for a generic theory, which may (not) have a Lagrangian description, no complete set of tests
exists that would rule it consistent; one usually has to work on a case by case basis, searching for inconsistencies
in the spectrum and studying its higher point functions.

Gauge/gravity duality~\cite{Maldacena:1997re,Gubser:1998bc,Witten:1998qj} can be a useful tool for determining
consistency constraints on the dynamics of quantum field theories, since it allows one to explore otherwise
unaccessible corners of the parameter space of the theory. As it often turns out, some pathological features are made more
manifest in one side of the duality or the other.

Consider for example the case of partially massless fields in
AdS~\cite{PM1,PM2,PM3}. Although they did not appear to spoil the consistency of the bulk theory at the classical level,
it was later shown~\cite{Dolan} that they correspond to operators in the dual boundary conformal field theory (CFT)
whose descendants create negative-norm states. Another example comes in the context of ghost-free higher derivative
gravitational theories, {\it i.e.}, Lanczos-Lovelock gravities~\cite{Lovelock:1971yv}
obtained by adding Euler density terms in the Einstein-Hilbert Lagrangian. The simplest such theory
is Gauss-Bonnet gravity in five dimensions, for which it was shown in~\cite{Myers,Buchel:2009tt,Hofman:2009ug}
that unless the Gauss-Bonnet coupling is appropriately bounded, the dual CFT will violate causality.
These results were subsequently generalized to arbitrary dimensions, Lovelock terms and other
higher derivative theories~\cite{GB,Lovelock}. More importantly, they were understood purely in the context of field
theory~\cite{Hofman:2009ug,Hofman:2008ar,Kulaxizi:2010jt} as necessary constraints imposed by unitarity.
In all the above examples, there seemed to be no inconsistency in the bulk theory, at least at the classical level;
it is the study of the boundary theory that reveals a pathology, whose cure, in turn, demands that the bulk dual be constrained.

%Gauge/gravity correspondence can be useful in understanding the consistency of a theory.
%Often, as it turns out,
%some pathological features are manifest only in one side of the duality or the other.
%In Gauss-Bonnet gravity, for
%example, the higher-derivative Gauss-Bonnet coupling must have an upper bound, because otherwise there %is violation
%of microcausality in the dual conformal field theory (CFT)~\cite{Myers,ManuelaGB}. Similarly, the parameters %in Lovelock
%gravities are constrained by causality considerations of the boundary theory~\cite{Lovelock}. Another %example is that
%of partially massless fields in AdS, for which negative-norm states show up in the boundary %CFT~\cite{Dolan}.

In this paper we will see that the AdS/CFT correspondence poses non-trivial constraints on the
electromagnetic (EM) and gravitational couplings of a charged massive spin-1 particle in the bulk.
These constraints presumably correspond to unitarity restrictions on the parameters which determine
certain correlation functions in the dual CFT.
We will consider a massive vector boson in the AdS Reissner-Nordstr\"om background.
In the dual CFT this amounts to considering the dynamics of a gauge-invariant spin-1 operator
at finite charge density, whose conformal dimension $\Delta$ is related to its mass in AdS.
We will find that the EM and gravitational couplings are constrained by consistency requirements
of the dual CFT. In the given background, it is a great convenience that cubic couplings
can be investigated at the level of a quadratic Lagrangian for the spin-1 field.  From the CFT point
of view it means that information about certain 3- and higher-point functions is contained in 2-point
functions at finite charge density.

To be more specific about the couplings we will be considering, let us note that if the EM interactions preserve Lorentz,
parity and time-reversal symmetries, a spin-1 particle of mass $m$ may possess a charge $q$, a magnetic dipole moment $\mu_m$,
and an electric quadrupole moment $Q_e$\,\footnote{In general a spin-$s$ particle can have $2s+1$ intrinsic multipole moments,
c.f.~\cite{Lorce} and references therein.}. The \emph{classical} electrodynamics is consistent for an arbitrary dipole
moment~\cite{Aronson,VZ}. But once this is chosen, causal propagation demands that the quadrupole moment be fixed~\cite{VZ}
such that one obtains (from~\cite{BH}, for example): \beq \mu_m=\frac{gq}{2m},\qquad Q_e=\frac{(1-g)q}{m^2},\eeq{EMmoments}
where $g$ is called the gyromagnetic ratio or $g$-factor, whose value is arbitrary in the classical theory. \emph{Quantum}
consistency of the theory, however, requires that the bare $g$-factor be fixed as well. Indeed, perturbative renormalizability
and tree-level unitarity constraints~\cite{Tiktopoulos}
uniquely give the ``natural" value, $g=2$, which is precisely the tree-level prediction of the standard model for the $W$
boson~\cite{SM,Kim}\,\footnote{The same values--$\mu_m=q/m$ and $Q_e=-q/m^2$--result from the Drell-Hearn-Gerasimov sum
rule~\cite{BH,Kim} and also from the requirement of (lightcone) helicity-preserving scattering amplitudes~\cite{Lorce,Kim}.}.
It turns out that $g=2$ is the ``preferred" tree-level value for \emph{all} spin~\cite{Weinberg,g=2,EM-SUSY}, and that open-string
theory predicts the same universal value as well~\cite{g=2,Strings}. Yet, the spin-1 case is exceptional and more interesting
in that for higher spins already the classical theory itself exhibits pathologies, e.g. superluminal
propagation~\cite{VZ,VZ1,Misc,DW}, which are (partially) remedied by fixing the $g$-factor~\cite{Strings,PR2}.

On the other hand, a spin-1 particle can have a gravitational quadrupole moment\,\footnote{Note that a gravitational dipole
term is not physically meaningful~\cite{GR-SUSY}.}, and associated with this coupling is the gravimagnetic ratio or
$h$-factor~\cite{GR-SUSY,Khriplovich}$-$the gravitational analog of the EM $g$-factor. The only constraint on $h$ known
to date comes in the presence of unbroken supersummetry, which demands $h=1$ for spin 1, and suggests the same value
for all higher spins~\cite{GR-SUSY}. In a non-supersymmetric theory, however, tree-level unitarity considerations
cannot single out any preferred value for the spin-1 $h$-factor\,\footnote{It does give the ``natural" value $h=1$ only for
spin $s>2$~\cite{h=1}.}. To spell out how the $g$-factor and the $h$-factor appear in the cubic interactions in a Lagrangian
describing a vector field $W_\mu$ coupled to EM and gravity, we write
\beq \mathcal{L}=-|\nabla_\mu W_\nu|^2+|\nabla_\mu W^\mu|^2-m^2W^\ast_\mu W^\mu
+iqgF^{\mu\nu}W^\ast_\mu W_\nu-hR^{\mu\nu}W^\ast_\mu W_\nu.\eeq{L0} Note that the minimal coupling prescription is ambiguous
because covariant derivatives do not commute: $[\nabla_\mu,\nabla_\nu]W^\alpha=iqF_{\mu\nu}W^\alpha+R^\alpha_{~\beta\mu\nu}W^\beta$;
the magnetic dipole and gravitational quadrupole couplings are therefore rather a consequence of this ambiguity.

In this article we will use holography to constrain these couplings. To this end, we analyze the equation of motions (EoM)
following from~(\ref{L0}) in the asymptotically AdS Reissner-Nordstr\"om background. In Section~\ref{sec:Background}
we review the details of the background, and in Section~\ref{sec:Fluctuations} we proceed to derive the spin-1 EoMs
which in a certain region of the parameter space can
%become relatively simple in a certain region of the parameter space where it is possible to find an explicit solution
be solved via the WKB method. This allows us to analytically determine the group velocity of the modes coupled to the
dual operator ${\mathcal O}_\mu$. Requiring these CFT modes not to propagate superluminally, gives constraints
on both $h$ and $g$ that we derive in Section~\ref{sec:Constranits}. Finally, in Section~\ref{sec:Conclusions} we clarify
and discuss the implications of our results and mention some interesting future directions.

%%%%%%%%%%%%%%%%%%%%%%%%%%%%%%%%%%%%%%%%%%%%%%%%%%%%%%%%%%%%%%
\section{Massive Spin 1: Holographic Analysis}\label{sec:Main}
%%%%%%%%%%%%%%%%%%%%%%%%%%%%%%%%%%%%%%%%%%%%%%%%%%%%%%%%%%%%%%

Consider a $d$-dimensional CFT with a global $U(1)$ symmetry
which has a dual description in terms of Einstein-Hilbert gravity in $d+1$-dimensional AdS.
The conserved current $J_\mu$ associated to this symmetry in the CFT is
mapped by the gauge/gravity duality map, to a $U(1)$ gauge field in AdS. At finite
charge density, {\it i.e.} $\langle J_t\rangle\neq 0$, the CFT is described by a charged black hole
in AdS. Let us further assume that the CFT contains at least one gauge invariant vector operator
${\mathcal O}_\mu$ of conformal dimension $\Delta>d-1$ which is charged under the $U(1)$ global symmetry
of the theory. It is this operator that is dual to a spin-1 field in the bulk of AdS, whose
dynamics is described by Lagrangian~(\ref{L0}).

%%%%%%%%%%%%%%%%%%%%%%%%%%%%%%%%%%%%%%%%%%%%%%%%%%%%%%%%%%%%%%%%%%%%%%
\subsection{AdS Reissner-Nordstr\"om Background}\label{sec:Background}
%%%%%%%%%%%%%%%%%%%%%%%%%%%%%%%%%%%%%%%%%%%%%%%%%%%%%%%%%%%%%%%%%%%%%%

Let us start by reviewing the charged AdS black hole geometry we will use.
The action for a photon field $A_\mu$ coupled to gravity in $\text{AdS}_{d+1}$ we consider is
\beq S={1\over 2 \kappa^2}\int d^{d+1}x \sqrt{-g}\left[R+{d(d-1)\over L^2}-{L^2\over g_F^2}F_{\mu\nu}F^{\mu\nu}\right],\eeq{s1}
where $g_F^2$ is the effective dimensionless gauge coupling, and $L$ is the curvature radius of AdS, which we henceforth set to
unity. The resulting EoMs are \beq G^{\mu\nu}-\tfrac{1}{2}\,d(d-1)g^{\mu\nu}=-\tfrac{2}{g_F^2}\left[
F^{\mu\rho}F_\rho^{~\nu}-\tfrac{1}{4}g^{\mu\nu}\text{Tr}(F^2)\right],\qquad \nabla_\mu F^{\mu\nu}=0,\eeq{eom}
which admit the charged black hole solution~\cite{AdSRN}: \beq ds^2=r^2\left[-f(r)dt^2+d\vec{x}^2\right]+{1\over r^2}{dr^2\over
f(r)},\qquad A_\sigma=\mu\left(1-{r_0^{d-2}\over r^{d-2}}\right)\delta_\sigma^t,\eeq{solution} where $f(r)$ depends on the
mass $M$ and the charge $Q$ of the black hole as \beq f(r)=1+{Q^2\over r^{2d-2}}-{M\over r^d},\eeq{s2} and the horizon radius
$r_0$ is determined by the largest positive root of $f(r)$, with \beq r_0^{2d-2}-Mr_0^{d-2}+Q^2=0.\eeq{s3} The chemical potential
$\mu$ and temperature $T$ of the dual boundary theory are given by \beq \mu=\sqrt{d-1\over 2(d-2)}\,{g_F Q\over r_0^{d-2}},\qquad T=\frac{dr_0}{4\pi}\left[1-\left(\frac{d-2}{d}\right)\frac{Q^2}{r_0^{2d-2}}\right].\eeq{s4}

We are interested in the zero temperature limit of the black hole, in which case $M$ and $Q$ become related to the horizon radius
as \beq M=\left(\frac{d-1}{d-2}\right)2r_0^d,\qquad Q=\sqrt{\frac{d}{d-2}}\,r_0^{d-1},\eeq{s5} so that the solution reduces to
\beq f(r)=1+{d\over d-2}\left({r_0\over r}\right)^{2d-2}-{2(d-1)\over d-2}\left({r_0\over r}\right)^d,\qquad \mu=\sqrt{\tfrac{1}
{2}d(d-1)}\,\frac{g_Fr_0}{d-2}.\eeq{s6} The AdS Reissner-Nordstr\"om background is this dynamical Maxwell-Einstein background
that satisfies the EoMs~(\ref{eom}) and is described by Eqs.~(\ref{solution}) and ~(\ref{s6}).

%%%%%%%%%%%%%%%%%%%%%%%%%%%%%%%%%%%%%%%%%%%%%%%%%%%%%%%%%%%%%%%%%%%%%%%%
\subsection{Spin-1 Fluctuations \& WKB Analysis}\label{sec:Fluctuations}
%%%%%%%%%%%%%%%%%%%%%%%%%%%%%%%%%%%%%%%%%%%%%%%%%%%%%%%%%%%%%%%%%%%%%%%%

The dynamics of the probe spin-1 field is
found by varying of the action~(\ref{L0}), which gives \beq \left(\nabla^2-m^2\right)W_\mu-\nabla_\mu\left(\nabla\cdot
W\right)+iq\left(2+\delta g\right)F_{\mu\nu}W^\nu-\left(1+\delta h\right)R_{\mu\nu}W^\nu=0,\eeq{eom1} where, for future
convenience, we have defined \beq \delta g\equiv g-2,\qquad \delta h\equiv h-1.\eeq{deltas} Another useful quantity
is the effective mass in AdS, \beq \mathfrak{m}^2~\equiv~m^2-d\,\delta h~\geq~0,\eeq{adsmass} which must be
non-negative, as we show in Appendix~\ref{sec:BF}. Notice that the second time derivative of $W_t$ never appears in the
EoM~(\ref{eom1}), so that one of the components of $W_\mu$ is non-dynamical as expected.

In the following we restrict ourselves to $d=4$, so that we will have a 5D bulk.
We would like to examine the 2-point function of the boundary operator
dual to $W_\mu$.
Rotational invariance allows us to consider small perturbations of the form $W_\mu(r,t,x_3)$, which
can be Fourier transformed as
\beq  W_\mu(r,t,x_3)=\int \frac{d\omega d k}{(2\pi)^2}\,{\hat W}_\mu(r)\,e^{i(kx_3-\omega t)}.\eeq{ftW}
From the point of view of the dual boundary theory we distinguish $\hat W_\mu(r)$ into transverse ($\mu=1,2$) and longitudinal
($\mu=0,3$) perturbations. Substituting~(\ref{ftW}) into Eq.~(\ref{eom1}) one observes, not surprisingly, that $\hat W_r$ is
not a dynamical field; it can be completely determined from the longitudinal modes via its EoM:
\beq {\hat W_r}(r)=\frac{ i \left(\omega-qA_t\right)W'_t+k f W'_3+(1+\delta g)q A'_t W_t}{(\omega-q A_t)^2
- f\left[k^2+\mathfrak{m}^2-\tfrac{1}{2}\delta{h}\,r^2 \left( 8 (f-1) +7 r f'+r^2 f'' \right)\right]}\,.\eeq{Wrsol}
On the other hand, the EoMs for the transverse modes completely decouple from the rest and will not be of interest to us here.

To study the longitudinal modes it is convenient to define a new set of fields:
\beq \begin{split} \mathcal{E}_{1}(r)&\equiv k {\hat W_t}(r)+\left[\omega-q A_t(r)\right]{\hat W_3}(r),\\
\mathcal{E}_{2}(r)&\equiv k^{-1}\left[\omega -q A_t(r)\right]{\hat W_t}(r)+ f(r) {\hat W_3}(r).\end{split}\eeq{EFdef}
%This field redefinition guarantees that the boundary 2-point functions of the operators
%dual to the fields $\mathcal{E}_{\pm}$ are completely decoupled (see Appendix~\ref{sec:EoM}).
%This can be eaisly verified by considering the boundary term of~(\ref{L1})
$\mathcal{E}_{i}(r)$, with $i=1,2$, satisfy a system of second order coupled differential equations. The equations
simplify considerably in a suitable limit of large frequency and momentum where they can be solved using the WKB
approximation. To be specific, let us define the following parameters (recall that we have set $L=1$)
\beq \tilde{\omega}=\frac{\omega}{r_0}\,,\qquad \tilde{k}=\frac{k}{r_0}\,,\qquad \tilde{\mu}=\frac{q\mu}{r_0}\,,\eeq{uvdef}
and a new radial variable $z\equiv \frac{r}{r_0}$\,, and take the following limit
\beq \tilde{k} z\gg1 \,\,\&\,\,  \tilde{\mu}z\gg1,\,\, \text{with}\,\, u\equiv \frac{\tilde\omega}{\tilde{k}}=\text{fixed}\,\,\&\,\,
v\equiv\frac{\tilde{\mu}}{\tilde{k}}=\sqrt{\frac{3}{2}}\,\frac{qg_F}{\tilde k}=\text{fixed}.\eeq{limita}
%Comment about Swinger pair production in the bulk.
In this limit, the EoMs for the modes $\mathcal E_i$ reduce to
\beq \begin{split} \mathcal{E}_{1}''+ \tilde{k} a(z) \mathcal{E}_{2}'+\tilde{k}^2 b(z) \mathcal{E}_{1} &=0,\\
\mathcal{E}_{2}''+\tilde{k} c(z)\mathcal{E}_{1}'+\tilde{k}^2 d(z) \mathcal{E}_{2}&=0,\end{split}\eeq{eqmla}
where the explicit form of the functions $a(z)$, $b(z)$, $c(z)$, $d(z)$ is given in Appendix~\ref{sec:EoM}.
For now let us just point out that $a(z),\, c(z)$ are proportional to $v \delta g$ so that
Eqs.~(\ref{eqmla}) decouple when $\delta g$ vanishes.
Note that in~(\ref{limita}) we have not scaled the effective mass $\mathfrak{m}$ with $\tilde{k}$.
Had we done so, the equations would have decoupled and the couplings $\delta g,\, \delta h$ would have been effectively
scaled to zero. This is why we resort to~(\ref{limita}). By taking this limit we focus on a regime in
the parameter space where instabilities due to Schwinger pair production and/or condensation of the spin-1 field
are more likely to occur. However, this will not seriously concern us here given that we will be interested in examining the near boundary region.
%\footnote{A sufficient condition to avoid pair-production is $\frac{q F_{rt}}{m^2}<<1$. }.
%This would also happen, {\it i.e.} the decoupling of
%the equations, had we also scaled the effective mass in AdS $\mathfrak{m}$ with $\tiled{k}$. In particular,
%in that case
%Note also, that $d(z)$ depends on all parameters of the theory, {\it{i.e.}}, $\delta g,v,u,m,\alpha,\beta$
%while $b(z)$ is only a function of $u$.

We proceed to solve~(\ref{eqmla}) with the WKB method which can be easily extended to a system of coupled second order
differential equations. A useful reference we will follow here is~\cite{wkb}.
%Here we will present only the basic ideas relevant for our purpose. The interested reader should
%consult~\cite{wkb} and references therein.
%We are searching for inconsistencies in the dual, boundary theory, when $\delta g\neq 0$.
The starting point is to consider the ansatz
\beq \mathcal{E}_{i}(z)=e^{i \tilde{k} S_{i}(z)},\qquad i=1,2,\eeq{ansatzepm}
where $S_{i}(z)$ has an expansion in negative powers of $\tilde{k}$
as: $S_{i}\equiv S_{i}^{(0)}+\tilde{k}^{-1}S_{i}^{(1)}+\cdots$.
Note that the standard amplitude factor for the WKB solution of a single mode
is hidden in the first order term $S^{(1)}_{i}$.
%whereas the leading WKB phase factor is $S^{(0)}$ with ..
Substituting~(\ref{ansatzepm}) into~(\ref{eqmla}) one finds that a sensible
solution requires $S^{(0)}_{i}=S^{(0)}$, {\it{i.e.}}, that $S^{(0)}$ be independent from the mode.
Subsequent examination of the leading term in $\tilde{k}$ yields
\beq \begin{pmatrix}
-p^2+b(z) & i a(z)p(z)\\
i c(z)p(z) & -p^2(z)+d(z)
\end{pmatrix}
\begin{pmatrix}
e^{i S^{(1)}_{1}}\\
e^{i S^{(1)}_{2}}
\end{pmatrix}=0, \eeq{eqsone}
where $p(z)$ is defined as $p(z)\equiv\frac{d S^{(0)}}{dz}$\,, following standard conventions. Existence of solution for
Eq.~(\ref{eqsone}) requires that the determinant of the $2\times 2$ matrix vanish, {\it {i.e.}},
\beq {\mathbb G}\equiv \begin{vmatrix}
-p^2+b(z) & i a(z)p(z)\\
i c(z)p(z) & -p^2(z)+d(z)
\end{vmatrix}=0.
%\Rightarrow p^4(z)+p^2(z)\left(a c-b-d\right)+b d=0
\eeq{Gdef}
Eq.~(\ref{Gdef}) yields two distinct solutions for $p^2$, which we denote as $p^2_\pm(z)$.
In a similar manner one can determine $S^{(1)}_{i}$. The leading order WKB solutions for the $\mathcal E_i$'s
turn out to be linear combinations of the following modes~\cite{wkb}:
\beq \mathcal{E}_{\pm}\simeq \mathcal{A}_{\pm}(z) e^{i \tilde{k} \int p_{\pm}(z)},\eeq{WKBSol}
where $\mathcal{A}_{\pm}(z)$ is inversely proportional to $\sqrt{\frac{\partial {\mathbb G}}{\partial p}}$\,.

%%%%%%REMOVE FROM EQUATION BUT COMMENT ON THE PHASE DIFFERENCE OF THE MODES%%%%%

%\beq \begin{pmatrix}
%\mathcal{E}_1(z) \\
%\mathcal{E}_2(z)
%\end{pmatrix} \simeq \begin{pmatrix}
%e^{i \tilde{k}\int p_1(z)} \\
%\frac{-p^2(z)+b(z)}{i a(z) p(z)}\, e^{i \tilde{k}\int p_2(z)}
%\end{pmatrix} \eeq{EFWKBsol}
%where $p_1(z)$ and $p_2(z)$ are the two distinct roots of the equation
%defined through the determinant of the coefficients of~(\ref{eqmla}),

In this work we are interested in solutions of fixed phase velocity (for the dual field theory mode). We
recall that the boundary frequency $\omega$ is shifted by $q\mu$ in the Reissner-Nordstr\"om background
(see for example~\cite{Faulkner:2009wj})
and thus the phase velocity is $u-v$. In the language of quantum mechanics and the WKB approximation,
the square of the phase velocity plays the role of the energy of the system and will henceforth be
denoted as $E$.
Solving~(\ref{Gdef}) with the help of the expressions in Appendix~\ref{sec:EoM} yields
\beq p^2_{\pm}=\frac{1}{z^4 f^2(z)}\left[\left(\sqrt{E_{\pm}}+\frac{v}{z^2}\right)^2- V_{\pm} \right],\eeq{pexp}
where
\beq
V_{\pm}(z)=\frac{f(z)}{\mathfrak{m}^2z^6-8\delta{h}}
\left[H(z)+\delta g^2 v^2 z^2\mp 2\sqrt{\delta g^2 v^2 z^2 H(z)+K^2(z)}\right],
\eeq{Vdef}
and $H(z)$, $K(z)$ are defined as follows
\beq H(z)=\delta{g}\left(2+\delta g\right)v^2z^2+\mathfrak{m}^2 z^6-2\delta{h},\qquad
K(z)=\delta{g}\,v^2 z^2+3\delta{h}.\eeq{HKdef}
% eq.(\ref{pexp}) becomes $p^2\sim z^{-4} \left(E-1\right)$. To avoid
%potential instabilities we thus search for solutions with "energy" $E\geq 1$.

The next step is to investigate possible turning points. It is shown in~\cite{wkb} that for a system of
coupled differential equations, turning points satisfy
\beq \left.\frac{\partial {\mathbb G}}{\partial p}\right |_{z_t}=0\qquad\Rightarrow\qquad p(z_t)=0 \quad\text{or}
\quad p(z_t)=-\tfrac{1}{2}(ac-b-d).\eeq{tpdef}
The first solution, $p=0$, is familiar from the study of the single channel WKB.
It corresponds to the point $z=z_t$ where $p(z)$ becomes imaginary.
The other solution, $p(z)=-\tfrac{1}{2}(ac-b-d)$, is a feature of the coupled WKB system and
corresponds to the point where $p_{+}(z)$ and $p_{-}(z)$ coalesce.
It is possible to show that the second class of turning point does not exist in this case (the reader
can refer to Appendix~\ref{sec:TP} for a proof). The same is not true, however, for turning points of the first class.
Unless the bulk couplings $(g, h)$ are appropriately constrained, special
points $z_t>1$ exist in the bulk for which $p^2(z)$ changes sign.

%In what follows we will systematically discuss the conditions on  under which such turning points exist.
%Before going into this detailed analysis, it is instructive to assume their existence and examine
%its implications.

For turning points of the first class, the standard approach for matching the solutions can be employed.
Treating the boundary at $z=\infty$ as an infinite wall yields the following quantization condition
\beq \tilde{k}\int_{z_t}^\infty p(z)+\mathcal{O}(\tilde{k}^0)=\pi \left(n \pm\tfrac{1}{4}\right),\qquad
n=1,2,...\,...\,.\eeq{qca}
Eq.~(\ref{qca}) allows one to compute the group velocity of the modes coupled to
the dual vector operator in the limit of~(\ref{limita}). The result is
\beq u_g\equiv \frac{d\omega}{d k}-v=\frac{\int u\frac{\partial p}{\partial u}+v \frac{\partial p}{\partial v} -p}
{\int \frac{\partial p }{\partial u} }-v\simeq \sqrt{E} \left[1+\frac{v}{\sqrt{E}}
\frac{\left( \frac{\partial p^2}{\partial v} \right)_u } {\left( \frac{\partial p^2}{\partial u} \right)_v}\right]_{z=z_t},\,\eeq{ug}
where the subscript in the parenthesis must be kept fixed under differentiation, while the bracket
is evaluated at the turning point $z=z_t$ corresponding to the maximum value of the energy $E$.
To derive~(\ref{ug}), we note that the integrands are strongly peaked around the turning
point, where $\left(\frac{\partial p}{\partial {\mathbb G}}\right)$ diverges. To see this it is convenient to
express the derivatives in the integrand in terms of $\left(\frac{\partial p}{\partial {\mathbb G}}\right)$ and
subsequently show that both $\left(\frac{\partial {\mathbb G}}{\partial u}\right)$ and
$\left(\frac{\partial {\mathbb G}}{\partial v}\right)$ are non-vanishing at the turning point (see Appendix~\ref{sec:TP}).
Using a standard identity from partial differentiation one can express the group velocity in a simpler form as
\beq u_g^2\simeq E\left[1-\frac{v}{\sqrt{E}}\left(\frac{\partial u}{\partial v}\right)_{p}\right]^2_{z=z_t} \, .\eeq{ug2}
%where we additionally used the defining property of the turning point $p(z_t)=0$.

The existence of a turning point in the WKB language, observed in~\cite{Myers} in the context of Gauss-Bonnet
gravity in AdS, is linked to causality violation of the dual theory. This has by now been confirmed
in a number of higher derivative gravitational theories~\cite{GB,Lovelock}. We will shortly see that
the same is true in our case; unless specific constraints are imposed on the bulk coupling parameters
$(g, h)$, the dual CFT will be inconsistent.

%%%%%%%%%%%%%%%%%%%%%%%%%%%%%%%%%%%%%%%%%%%%%%%%%%%%%%%%%%%%%%%%%%%
\subsection{Constraints on Spin-1 Couplings}\label{sec:Constranits}
%%%%%%%%%%%%%%%%%%%%%%%%%%%%%%%%%%%%%%%%%%%%%%%%%%%%%%%%%%%%%%%%%%%

First of all let us note that the potential $V_{\pm}(z)$, given by Eq.~(\ref{Vdef}), vanishes at the horizon $z=1$
and tends to unity at the boundary $z=\infty$. A classical turning point $z_t$ will exist if $V_{\pm}(z_t)>1$.
This will pertain to the ``energy" value $E_\pm=V_\pm(z_t)$, which is of course greater than one. However, no turning
points will exist if $V_{\pm}(z)$ never exceeds unity.

Then Eq.~(\ref{ug2}) implies that for a neutral spin-1 field, and in general for $v\ll1$ as we will see,
%\beq \frac{v}{\sqrt{E}}\ll1~\Leftrightarrow~\frac{q\mu}{\omega-q\mu}\ll1,\eeq{smallv}
the existence of a turning point leads to causality violation, because the group velocity (along with the phase velocity)
will exceed unity. A \emph{necessary} condition to avoid this is that turning points do not exist. This requirement
can be fulfilled by appropriately constraining the parameter space of the bulk couplings.

Let us first examine the simpler case of a neutral spin-1 field, {\it{i.e.}}, $v=0$.
The potential $V_{\pm}$ defined in Eq.~(\ref{Vdef}) reduces to\,\footnote{To avoid dealing with
ambiguities related to the sign of the square root it is perhaps best to work out the WKB potential
directly from the Eqs.~(\ref{eqmla}), which in this case decouple.}
\beq V_{\pm}=\left(1-\frac{3}{z^4}+\frac{2}{z^6}\right)\left[\frac{\mathfrak m^2 z^6-2(1\pm3)\delta h}
{\mathfrak m^2 z^6-8\delta h}\right],\eeq{Va}
From Eq.~(\ref{Va}) it is easy to see that $V_{+}$ never exceeds unity for any $\delta h$.
On the other hand, $V_{-}$ remains bounded if and only if the
denominator in Eq.~(\ref{Va}) is nowhere vanishing in the bulk\,\footnote{Here the potential is not
bounded from above and diverges fast as the denominator vanishes. While the quantization condition~(\ref{qca})
must be slightly modified, the group velocity~(\ref{ug2}) is still correct.}. This requirement
translates to the following constraint for the effective mass: $\mathfrak m^2>8\delta h$.
It is easy to see that negative values of $\delta h$ do not lead to causality violation. However, for
some negative values of $\delta h$ the potential $V_{-}$ becomes negative as well. Although not related to
causality violation, negative values of the potential have been shown~\cite{Myers:2007we,Stability} to
signal instabilities. Avoiding them requires $\mathfrak m^2\geq-4\delta h>0$. Combining these requirements
leads to\,\footnote{Considering non-extremal RN-black hole does not provide additional constraints;
one can easily determine the fluctuation equations at finite but non-zero temperature in the large momentum and
frequency limit to find that the form of the potential remains essentially unchanged.}
\beq -\tfrac{1}{4}\mathfrak{m}^2~\leq~\delta{h}~<~\tfrac{1}{8}\mathfrak{m}^2.\eeq{Vacon}

It is interesting to consider the limit of vanishing mass when gauge symmetry is restored and there exists one (longitudinal) physical
degree of freedom in the bulk $-$ usually associated to $\mathcal{E}_{1}$\,. However, in the large frequency and momentum limit discussed
here both the $\mathcal{E}_{i}$'s are physical since they are simply proportional to one another.
This is because for parametrically large frequency and momentum the dual CFT is effectively at zero charge density.
As a result, Lorentz invariance is restored, $\omega\simeq k$ and $\mathcal{E}_{2}\sim {1\over k} \mathcal{E}_{1}$.
It is therefore natural to demand consistency of the theory for both modes when $\mathfrak{m}=0$.
Eq.~(\ref{Vacon}) implies in this case that $\delta h$ must vanish.

%\begin{figure}
%\begin{center}
%\includegraphics[width=10cm,height=7cm,keepaspectratio]{veffplus.jpg}
%\caption{\small This figure shows $V^{\text{eff}}_{+}(z)$ as a function of the radial variable $z$ for fixed $\mathfrak{m}^2=9,\,\delta h=-2,\,v=3$
%and four different values of $\delta g=4,\,6,\,8,\,10$. As $\delta g$ increases the potential increases faster towards the boundary
%where it reaches its maximum value $V_{+}^{\text{eff}}(z=\infty)=1$.}
%\end{center}
%\end{figure}

From the point of view of the bulk requiring consistency of the theory for arbitrary mass seems quite natural.
However, for the dual CFT the bulk mass is related to the conformal dimension $\Delta$ of the dual operator through
$\mathfrak{m}^2=\Delta \left(\Delta-3\right)$. Unitarity of the CFT requires $\Delta\geq 3$ and the inequality is saturated
for a conserved spin-1 current dual to a gauge field in the bulk. Clearly, $\Delta$ is not a continuous variable, in the sense that
a certain CFT does not necessarily contain spin-1 operators of all possible conformal dimensions.
We are thus hesitant to set $\delta h=0$.
Nevertheless, to the best of our knowledge the constraint~(\ref{Vacon}) on the spin-1 gravitational quadrupole moment
is completely new  for a non-supersymmetric theory.

%\begin{figure}
%\begin{center}
%\includegraphics[width=20cm,height=7cm,keepaspectratio]{veffminus.jpg}
%\caption{\small Here we see a plot of $V_{-}^{\text{eff}}(z)$ in terms of $z$ for $\mathfrak{m}^2=9,\,v=3,\,\delta h=-2$,
%and for three different values of $\delta g=2,\,3,\,3.5$. When $\delta g$ becomes greater than $\mathfrak{m}$ the potential
%attains a maximum value greater than one in the bulk.}
%\end{center}
%\end{figure}

Now let us consider the case of non-zero charge. We are interested in $v\ll1$ since we want to probe the properties of the boundary
theory close to the conformal point. From the CFT perspective we expect the constraint~(\ref{Vacon}) to be unaffected by a non-zero
$v$. This is because $\delta h$ corresponds to a parameter appearing in some CFT correlator, which of course is independent of the
global $U(1)$ charge of the dual vector operator. Thus in what follows we expect Eq.~(\ref{Vacon}) to be valid.
The effective potential $V^{\text{eff}}$ in this case is defined as
\beq V^{\text{eff}}_{\pm}(z)\equiv \left(\sqrt{V_{\pm}}-\frac{v}{z^2}\right)^2,\eeq{Veff}
such that at the turning point $z_t$, we have $E=V^{\text{eff}}_{\pm}(z_t)$. Note that
the ambiguity in the sign in front of $\sqrt{V_\pm}$ in principle leads to two distinct
potentials for the spin-1 field. It turns out, however, that consistency demands the
sign to be the same as that of the charge. Here we consider $v\geq 0$ and thus choose
the positive sign in~(\ref{Veff}).
\begin{figure}[ht]
\begin{minipage}[b]{0.5\linewidth}
\centering
\includegraphics[width=1\linewidth,height=.65\linewidth]{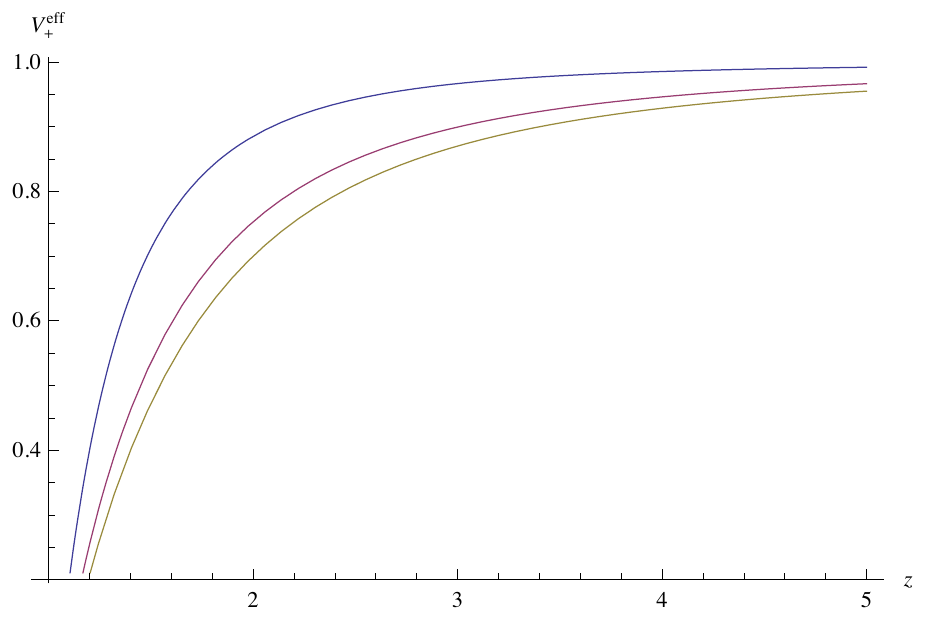}
\caption{\footnotesize This figure shows $V^{\text{eff}}_{+}(z)$ for fixed $v=0.10,\,\mathfrak{m}^2=1.00,\,\delta h=0.06$,
and three different values of $\delta g=0.50,\,7.00,\,10.00$. As $\delta g$ increases the potential increases faster towards the boundary
where it reaches a local maximum with $V_{+}^{\text{eff}}(z=\infty)=1$.}
\label{fig:figure1}
\end{minipage}
\hspace{0.5cm}
\begin{minipage}[b]{0.5\linewidth}
\centering
\includegraphics[width=1\linewidth,height=.65\linewidth]{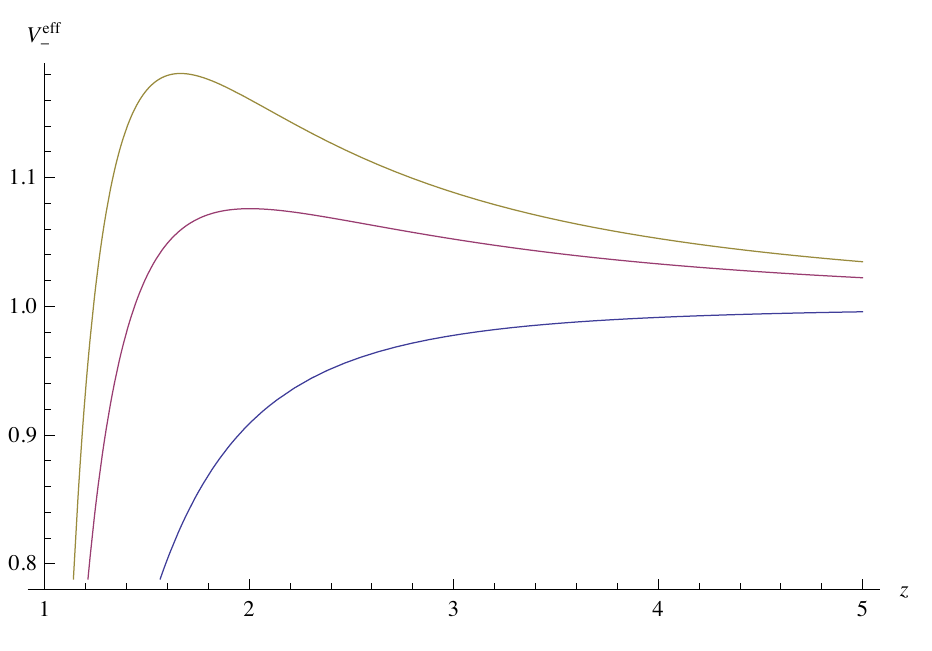}
\caption{\footnotesize Here we see a plot of $V_{-}^{\text{eff}}(z)$ versus $z$ for fixed $v=0.10,\,\mathfrak{m}^2=1.00,\,\delta h=0.06$,
and for $\delta g=0.50,\,7.00,\,10.00$. When $\delta g^2$ becomes greater than $\mathfrak{m}^2$ the potential attains a local maximum whose value exceeds unity.}
\label{fig:figure2}
\end{minipage}
\end{figure}

It is instructive to study how $V_\pm^{\text{eff}}$ change as $\delta g$ and $\delta h$ are varied.
A plot of the effective potential of each mode for various values of $\delta g$ where $\delta h$ and
$\mathfrak m$ are held fixed, is presented in Fig.\,1 and Fig.\,2\,\footnote{In the limit of zero charge
the results reduce to those of the previous paragraph.}. On the one hand, $V^{\text{eff}}_{+}$ smoothly
decreases except for a tiny region close to the horizon, as expected from~(\ref{Veff}). $V^{\text{eff}}_{-}$,
on the other hand, develops a local maximum with $\text{Sup}\left\{ V^{\text{eff}}_{-} \right\}>1$ for some values
of $\delta g$. In this case the existence of a turning point for $E>1$
is unambiguous and the potential supports metastable states whose group velocity may exceed unity.

To examine the conditions under which causality is violated let us first expand the effective potential near the boundary:
\beq
V^{\text{eff}}_{-} = 1+\frac{2 v}{z^2}\left[\frac{|\delta g|}{\sqrt{\mathfrak m^2}}-1\right]
-\frac{3}{z^4}\left[1-\frac{v^2}{3}\left(\left[\frac{|\delta g|}{\sqrt{\mathfrak m^2}}-1\right]^2+
\frac{\delta g^2}{\mathfrak m^2}+\frac{2\delta g}{\mathfrak m^2}\right)\right]+\mathcal{O}\left(\frac{1}{z^6}\right).
\eeq{Veffnb}

We see that whenever $\delta g^2>\mathfrak m^2$, even for $v\ll1$, the potential will develop a maximum. As long as the
coefficient of the $\mathcal{O}(v^2)$ term is not too large, the first three terms in the near-boundary expansion~(\ref{Veffnb})
suffice to give an estimation of the maximum of the potential. This in turn enables one to compute the group velocity~(\ref{ug2}),
which has the following power-series expansion in $v$:
\beq u_g=1+v-\frac{v^2}{6}\left(\frac{|\delta g|}{\sqrt{\mathfrak m^2}}-1\right)^2+\mathcal{O}(v^4).\eeq{ugvexpand}
It is clear that for $v\ll1$ the group velocity always exceeds unity for finite values of $\delta g^2/\mathfrak m^2$.
Causality violation can be avoided by forbidding the potential to develop a maximum with the condition:
\beq \delta g^2~\leq~\mathfrak{m}^2.\eeq{dgcon}

Note that if $v$ ceases to be small, the effective potential might develop a maximum even for values of $\delta g^2$ within the
bound~(\ref{dgcon}). This however does not necessarily imply causality violation since the group velocity may still remain
smaller than unity. In any case, analyzing this regime of $v$ is beyond the scope this article.

\section{Conclusions \& Outlook}\label{sec:Conclusions}
%%%%%%%%%%%%%%%%%%%%%%%%%%%%%%%%%%%%%%%%%%%%%%%%%%%%%%%

In this paper, we have shown that holography constrains the EM and gravitational couplings of a spin-1 field.
For a given mass, the otherwise undetermined $g$-factor and $h$-factor of the classical theory are allowed to take values
only in some given range.
%given respectively By Eqs.~(\ref{dgcon}) and ~(\ref{Vacon}).
The results followed from considering in AdS Reissner-Nordstr\"om background the dynamics of spin-1 fluctuations, whose EoMs
simplify in a certain region of the parameter space, so that one could employ the WKB method to construct explicit solutions.
For the modes coupled to the dual boundary operator, one could then calculate the group velocity, which become superluminal
unless the spin-1 couplings are constrained.

Field theoretically, the Lagrangian~(\ref{L0}) describes generic electromagnetic and gravitational couplings of
a massive vector boson. The bounds~(\ref{Vacon}) and~(\ref{dgcon}), however, are obtained for a very particular
setup $-$ when the spin 1 is a probe in a dynamical Maxwell-Einstein background with the AdS Reissner-Nordstr\"om
geometry, so that holography could be employed. The generic situation of an arbitrary background may not come with
a holographic interpretation, but is likely to call for sharper bounds. Since the couplings $q$, $g$, $h$ and the
original mass $m^2$ (which one takes to be positive) appearing in the Lagrangian~(\ref{L0}) are a priori independent
of one another, one expects the same for the bounds on these couplings. This possibility leads to the
``natural'' value of $g=2$ and to $h\leq1$.

It is quite possible that a more extensive holographic analysis, which takes into consideration other regions of the
parameter space, could yield stronger constraints. From the point of view of the CFT, the constraints obtained here on
$\delta g$ presumably correspond to unitarity constraints on the parameters that determine the 3-point function of two spin-1
operators and a conserved vector current (or stress energy tensor operator)~\cite{Hofman:2008ar,Kulaxizi:2010jt}.
After all, their holographic derivation is remarkably similar to that appearing in~\cite{Myers}.
The black hole background used in~\cite{Myers} to
provide a non-zero expectation value for the stress energy tensor operator, is replaced here by the Reissner-Nordstr\"om which provides
a non-zero charge density (or chemical potential). While Ref.~\cite{Myers} studied the 2-point
function of the stress energy tensor, in this work we consider the 2-point function of a non-conserved spin-1 operator.

Given this interpretation, one could try to reproduce our results in field theory by following the reasoning
of~\cite{Kulaxizi:2010jt}.  Consider for example the 2-point function of a spin-1 operator at finite charge density
$\rho\equiv \langle J_t\rangle$ with the help of the operator product expansion (ope),
\beq \langle \mathcal{O}^{}_\mu (-k)\mathcal{O}_{\nu}(k)\rangle =\cdots+\mathcal{A}_{\mu\nu}^{~~\;\sigma}(k)\langle
J_\sigma\rangle+\cdots\,,\eeq{ope} where the dots represent the contribution of other operators in the ope.
Symmetries imply that the ope coefficient $\mathcal{A}_{\mu\nu\sigma}$ can only depend on a few
(generically coupling constant dependent) parameters. Moreover it is possible to show that in the large frequency and large
momentum limit, {\it i.e.}, when ${\omega\over\rho}\gg 1$ and ${k\over\rho}\gg 1$ the term explicitly shown in~(\ref{ope}) provides
the leading contribution to the 2-point function. Positivity of the spectral density of the 2-point function will then necessarily
impose constraints on the parameters which appear in  $\mathcal{A}_{\mu\nu\sigma}(k)$.

However, there is an obvious issue with this line of reasoning. Unlike the thermal expectation value of the stress energy tensor
which is necessarily positive definite (since it is the energy of the system), the charge density expectation value can have
either sign. The same issue arises if we try to interpret these constraints as charge flux constraints in analogy with the
energy flux constraints of~\cite{Hofman:2008ar}.
The charge flux operator was already defined in~\cite{Hofman:2008ar} but to the best of our knowledge there is no
physical principle which would restrict its 1-point functions to be non-negative.
Perhaps the resolution of this puzzle lies in understanding the parameter-space region where the results are derived $-$ as one
can see from~(\ref{limita}), the chemical potential is scaled with the momentum in the large momentum region. It would be interesting
to investigate this point further and give a proper identification of the couplings $g$ and $h$ in terms of the dual CFT.

We emphasize again that the parameters which determine the 3-point function of two spin-1 operators and a conserved current are
in general 't Hooft coupling dependent\,\footnote{This is different from what happens to the parameters which specify the 3-point
function of the stress energy tensor.}. Hence if the cubic couplings discussed here correspond to these CFT parameters, Eqs.~(\ref{dgcon})
and~(\ref{Vacon}) would represent constrains at strong coupling. In this respect our results may have some predictive value, especially
since consistency of the dual CFT for spin-1 operators of arbitrary conformal dimension requires that $\delta g=0$\,\footnote{The case
of $\delta h$ is somewhat different. Thanks to field redefinitions, it is expected to be related to certain 4-point functions.}.
%Especially given that the
%conformal dimension of a non-conserved spin-1 operator latter generically depends on the 't Hooft
%coupling.

%Since bounds on the Gauss-Bonnet coupling constant in \cite{Myers} correspond  one would be tempted to think that
%eqs.~(\ref{dgcon}) corresponds to charge flux
%constraints. A charge flux operator can be defined in four spacetime dimensions in complete analogy with the
%energy flux operator as~\cite{Hofman:2008ar}
%\beq \mathcal{Q}(\overrightarrow{n})=\lim_{r\rightarrow\infty} r^2\int_{-\infty}^\infty dt n^i J_i(t,%\,r\overrightarrow{n})
%\eeq{cfdef}
%where $J$ is the current associated to a global $U(1)$ symmetry of the theory.

%analogy with what happens in [MaldacenaHofman]
%mass related to dimension of operator in the dual CFT. the only value consistent for operators
%of any dimensionality which satisfies the unitarity bound, is zero. Corrections from coupling may
%change Delta - then the theory would not remain unitary unless the scaling dimension was increasing.
%Similarly when there is a flow...?? dangerously irrelevant operators and spin-one...................... END MANUELA

In this work we have seen that for a bulk theory which is perfectly consistent classically, the boundary dual may be inconsistent.
Could it be that the CFT actually probes the quantum consistency of the theory in the bulk? An affirmative answer
might find justification in the fact that perturbative renormalizability of the spin-1 electrodynamics indeed
requires the $g$-factor to be constrained (to the bare value of 2)~\cite{Tiktopoulos}. Classically consistent
in $\text{AdS}_3$, some Vasiliev-like theories also seem to support this, since they are believed to lack a healthy
quantum description and the corresponding CFT is known to be non-unitary~\cite{Perlmutter}\,\footnote{We
thank R.~Gopakumar for clarifying this point.}.
One might therefore argue that the duality holds good if the full quantum theory in the bulk, at least in the weak
coupling regime, is well behaved. As we have already mentioned, weak coupling requires $g\approx2$ for all spin~\cite{Weinberg}.

%Another way of putting this viewpoint is that the duality itself
%holds given that the full quantum theory in the bulk is well behaved. On the other hand, one might also argue that
%the duality holds good with the bulk theory being weakly coupled. .

One might wonder what would be the consequences of other possible cubic couplings added to the Lagrangian~(\ref{L0}).
At the 2-derivative level, indeed, there is the non-minimal coupling to the scalar curvature: $RW^\ast_\mu W^\mu$.
This term redefines the effective mass, but has no contribution to the gravitational quadrupole~\cite{GR-SUSY,h=1}.
Inclusion of this term would not affect our analysis, and we have chosen to drop it for the sake of simplicity.
Among the possible higher-derivative cubic couplings the only one relevant for our analysis is the ad hoc 3-derivative EM
quadrupole term: $W^*_\mu\nabla^\mu F^{\nu\rho}\nabla_\nu W_\rho+\text{h.c.}$, which alters the value of $Q_e$ given
in Eq.~(\ref{EMmoments}). But such a term is inconsistent to begin with, since it afflicts the classical bulk theory with
superluminal modes~\cite{VZ}.

In principle, for particles of any spin one should be able to do similar analysis to constrain their EM and gravitational
couplings. Spin $\tfrac{1}{2}$ is particularly interesting in this respect because it has a striking similarity with spin 1:
the classical electrodynamics of either particle allows an arbitrary $g$-factor but quantum consistency (perturbative
renormalizability) of the theory requires in either case that $g=2$.
It was recently appreciated that cubic couplings for a spin-$\tfrac{1}{2}$ field show up naturally
in top-down AdS/CFT models~\cite{Ammon:2010pg,CT,DeWolfeab,MT}.
%in consistent truncations of eleven and ten-dimensional (IIB) supergravity to AdS over Sasaki-Einstein manifolds~\cite{CT,MT}.
In these class of models, the $g$-factor for given mass and charge of the fermion can take several discrete
values\,\footnote{We are thankful to M.~Taylor for discussions on this issue.}$^,$\footnote{This does not necessarily
contradict the perturbative results quoted above since low energy effective Lagrangians derived, {\it e.g.}, from consistent truncations
usually contain additional fields. These extra fields can make the theory well behaved away from the point $g=2$.
See also~\cite{Ammon:2010pg,spin1/2} for the effects of an arbitrary magnetic dipole term of spin-$\tfrac{1}{2}$ fields on holographic Fermi surfaces.}.
%Clearly, it is worth investigating this further.

One would also like to study higher spins, for which the bulk theory is generically fraught with inconsistencies
already at the classical level~\cite{VZ,VZ1,Misc,DW}. One might consider, for example, the next simple case of a spin-2 field
$\varphi_{\mu\nu}$, for which the $g$- and $h$-factors appear the Lagrangian in the following way.
\bea \mathcal L&=&-|\nabla_\mu \varphi_{\nu\rho}|^2+2|\nabla_\mu \varphi^{\mu\nu}|^2+|\nabla_\mu\varphi|^2
+\left(\varphi^\ast_{\mu\nu}\nabla^\mu\nabla^\nu \varphi+\text{c.c.}\right)-m^2
\left(|\varphi_{\mu\nu}|^2-|\varphi|^2\right)\nonumber\\&&+2iqgF^{\mu\nu}\varphi^\ast_{\mu\rho}\varphi_\nu^{\rho}
+2h\varphi^\ast_{\mu\nu}\left(R_{\mu\alpha\nu\beta}\varphi^{\alpha\beta}-R^{\mu\alpha}\varphi_\alpha^\nu\right)
+...\,,\eea{two1} where $\varphi^\mu_\mu\equiv\varphi$, and the ellipses stand for non-minimal terms that are not important
for our purpose\,\footnote{Consistency requires that the coefficients of such terms are not all
independent~\cite{PM3,d-wave}.}. Here we expect that $g$ and $h$ will be constrained as well, just as they are for the spin-1 field.
It is worth mentioning that recently a spin-2 action in $\text{AdS}_5$ was derived by consistent truncation of type IIB supergravity on
an Einstein-Sasaki manifold, where the gyromagnetic ratio does not take the natural value of 2~\cite{MT}.
It will be interesting to see what implications the pathologies of the {\it classical} theory in the bulk may have for the dual CFT.
We leave these as future work.

%%%%%%%%%%%%%%%%%%%%%%%%%%%%%
\subsection*{Acknowledgments}
%%%%%%%%%%%%%%%%%%%%%%%%%%%%%

We would like to thank R.~Argurio, A.~Parnachev and M.~Taylor for useful discussions. The work of MK is partially supported by the ERC
Advanced Grant ``SyDuGraM", by IISN-Belgium (convention 4.4514.08) and by the ``Communaut\'e Fran\c{c}aise de Belgique" through
the ARC program. RR is a Postdoctoral Fellow of the Fonds de la Recherche Scientifique-FNRS. His work is partially supported
by IISN-Belgium (conventions 4.4511.06 and 4.4514.08) and by the ``Communaut\'e Fran\c{c}aise de Belgique" through the ARC program.

\begin{appendix}
\numberwithin{equation}{section}

%%%%%%%%%%%%%%%%%%%%%%%%%%%%%%%%%%%%%%%%%%%%%%
\section{Stability Bound in AdS}\label{sec:BF}
%%%%%%%%%%%%%%%%%%%%%%%%%%%%%%%%%%%%%%%%%%%%%%

In pure $\text{AdS}_{d+1}$, let us consider a massive vector field that does not backreact on the geometry. The dynamical background
satisfies the EoM \beq R^{\mu\nu}=-dg^{\mu\nu},\eeq{BF0} which simplifies the last term in the spin-1 action~(\ref{L0}), and reduce it
to \beq \mathcal S = \int d^{d+1}x\, \sqrt{-g}\, \Big(-\tfrac{1}{2}\mathcal{F}_{\mu\nu}^*\mathcal{F}^{\mu\nu}
-\mathfrak m^2W_\mu^*W^\mu\Big),\eeq{BF1} where $\mathcal F_{\mu\nu}\equiv\nabla_\mu W_\nu-\nabla_\nu W_\mu$
is the spin-1 curvature, and $\mathfrak{m}^2\equiv m^2-d(h-1)$ is the effective mass in AdS. One can show that
\beq \mathfrak m^2~\geq~0,\eeq{BF2} if the energy functional for $W_\mu$ is required to be positive definite\,\footnote{This is in fact
the spin-1 counterpart of the BF bound for scalar fields in AdS~\cite{BFbound}.}. Apart from the fact that a generic non-zero value of
$h-1$ changes the effective mass, the proof is essentially given in Appendix B of~\cite{d-wave}, which we rephrase here for the
sake of self-containedness.

The stress-energy tensor for the spin-1 field is \beq \mathcal T_{\mu\nu} = \frac{2}{\sqrt{-g}}\,\frac{\delta\mathcal S}{\delta
g^{\mu\nu}}\,,\eeq{BF3} which, when integrated over a spacelike slice orthogonal to the timelike Killing vector $\xi^\mu = (1,0,\dots,0)$,
gives the energy functional \beq \mathcal E = \int d^dx\, \sqrt{-g}\,\mathcal T^{0\mu} \xi_\mu.\eeq{BF4} To compute this quantity,
one can choose the line element \beq ds^2=(1/\cosh^2\rho)(-dt^2 + d\rho^2 + \sinh^2\rho\, d\Omega_{d-1}),\eeq{BF5} and
distinguish the time component ($\mu = 0$) from the spatial ones ($\mu,\nu = i,j$). One finds that the energy functional~(\ref{BF4})
boils down to \beq \mathcal E = \int d^dx\, \tanh^{d-1}\rho\, \left[g^{ij}\mathcal F_{0i}^*\mathcal F_{0j} + \frac1{2\cosh^2\rho}\,
\mathcal F_{ij}^*\mathcal F^{ij}+\mathfrak m^2|W_0|^2+\frac{\mathfrak m^2}{\cosh^2\rho}\, W_i^*W^i \right].\eeq{BF6}

Note that $\mathcal F_{\mu\nu}=0$ is always a valid solution, even when $\mathfrak m^2\neq0$. To see this, one notices that the energy
functional~(\ref{BF6}) depends on neither $\partial_0 W_0$ nor $\partial_\rho W_\rho$. Therefore, at a given time one can construct modes
with only one non-zero component: $W_\rho(\rho)$, which vanishes at large $\rho$ arbitrarily fast and satisfies $\partial_0 W_\rho = 0$.
When $\mathfrak m^2\neq0$, there appears the constraint $\nabla\cdot W=0$, but it merely fixes the quantity $\partial_0 W_0$.

Given this, if we demand that $\mathcal E$ be positive definite, restrictions on $\mathfrak m^2$ follow immediately. When $\mathfrak m^2 >0$,
the energy functional~(\ref{BF6}) is manifestly positive definite. When $\mathfrak m^2=0$, solutions with $\mathcal F_{\mu\nu}=0$ have zero
energy, so that $\mathcal E$ is semi-positive definite. However, because energy is interpreted as the norm in the Hilbert space of states,
such solutions are unphysical and correspond to null states. If $\mathfrak m^2<0$, the energy functional can become negative, which is
obviously the case for solutions with $\mathcal F_{\mu\nu}=0$. Correspondingly, the Hilbert space is plagued with negative-norm states.
This sets the bound~(\ref{BF2}).

%%%%%%%%%%%%%%%%%%%%%%%%%%%%%%%%%%%%%%%%
\section{The Spin-1 EoMs}\label{sec:EoM}
%%%%%%%%%%%%%%%%%%%%%%%%%%%%%%%%%%%%%%%%

Under the limit~(\ref{limita}), the EoMs for the modes $\mathcal E_i$ reduce to~(\ref{eqmla}), which contain the functions
$a(z)$, $b(z)$, $c(z)$ and $d(z)$, whose explicit forms are given below.
\bea a(z)&=&2\delta g\,v z^3N_1^{-1}(z),\\ b(z)&=&\tfrac{z^2 N_1(z)}{\left(z^2-1\right)^4\left(z^2+2\right)^2}\,,\\
c(z)&=&2\delta g\,v z N_2^{-1}(z),\\ d(z)&=&-\tfrac{4 \delta g^2v^2z^6(z^2-1)^2(z^2+2)((u-v)z^2+v)^2+4 \delta g\,v^2z^4
(z^2-1)^2(z^2+2)N_1(z)+z^2N_1(z )N_3(z)}{\left(z^2-1\right)^3\left(z^2+2\right)\left(z^4+z^2-2\right)N_1(z)N_2(z)}\,,\eea{abcd}
where the functions $N_1(z)$, $N_2(z)$ and $N_3(z)$ are defined as
\bea
N_1(z)&=&z^6\left[(u-v)^2-1\right]+2z^4v(u-v)+z^2(v^2+3)-2,\nonumber\\
N_2(z)&=&\mathfrak{m}^2z^6-8\delta h,\\
N_3(z)&=&4\delta{h}\left[z^6\left(2(u-v)^2+1\right)+4z^4v(u-v)+z^2(2v^2-3)+2\right]-\mathfrak{m}^2z^6 N_1(z).\nonumber\eea{Ndef}

%%%%%%%%%%%%%%%%%%%%%%%%%%%%%%%%%%%%%%
\section{Turning Points}\label{sec:TP}
%%%%%%%%%%%%%%%%%%%%%%%%%%%%%%%%%%%%%%

First we show that there exist no turning points of the second class, defined in~(\ref{tpdef}).
In other words, we show that the polynomial below--the square root in Eq.~(\ref{Vdef})--has no real solution
for $z>1$. In the following, we set $z^2=y$ and define the polynomial in question
\beq Z(y)\equiv v^2\mathfrak{m}^2\delta g^2\left[y^4+{\left(1+\delta g\right)^2v^2\over \mathfrak{m}^2}\,
y^2+2\delta h{\left(3-\delta g\right)\over \delta g\,\mathfrak{m}^2}\,y+
{9\delta h^2\over \delta g^2v^2\mathfrak{m}^2}\right]
%\delta g^2 v^2 y \left[\delta{g}\left(2+\delta g\right)v^2 y+\mathfrak{m}^2 %y^3-2\delta{h}\right]+\left(\delta g\, v^2 y+3\delta h\right)^2
\eeq{Zdef}
Notice first that $Z(y)$ behaves like $y^4$ near the boundary and is non-negative at the horizon $y=1$.
For the allowed values of $\delta h$, {\it i.e.},
$-\tfrac{1}{4}\mathfrak{m}^2\leq \delta h<\tfrac{1}{8}\mathfrak{m}^2$, it is in fact possible to show that
$Z(y)$ is non-negative everywhere in the bulk for $y>1$. This implies that any real roots outside the horizon
will coincide with a point where $Z(y)$ has a minimum. In other words, any root should also be a root of the derivative
of the polynomial. From Descartes rule of signs, one sees that $Z(y)$ will generically have only complex roots.
It is only when the coefficient of the linear term in~(\ref{Zdef}) is negative and the following inequalities are
met, namely,
\beq \begin{split}
&\delta h>0\quad\&\quad 0>\delta g>{-3\delta h\,\over 2\mathfrak{m}^2-\delta h}\quad\&\quad  |v|<\sqrt{\frac{-2\delta g\,
\mathfrak{m}^2-\delta h\left(3-\delta g\right)}{\delta g\left(1+\delta g\right)^2}}\,,\\
&\delta h<0\quad\&\quad 0<\delta g<{-3\delta h\,\over 2\mathfrak{m}^2-\delta h}\quad\&\quad  |v|<\sqrt{\frac{-2\delta g\,
\mathfrak{m}^2-\delta h\left(3-\delta g\right)}{\delta g\left(1+\delta g\right)^2}}\,,\\
\end{split}
\eeq{con1}
may the polynomial have two real roots outside the horizon.
Since every such root of $Z(y)$ must also be a root of $Z'(y)$, and $Z'(y)$ is a cubic polynomial with
just one real root, the only possibility is that $Z(y)$ has a double real root. However, this is only true
if a relation exists between the coefficients of the polynomial which cannot be satisfied for generic
$\delta g\neq 0,\,v\neq 0$. As a result there are no second class turning points.

%It is now not so difficult to see that $Z(z)$ is a monotonic function of $z$. If it were not, then $Z'(z)$ would
%vanish for some $z_0>1$. However from~(\ref{Zdef}) we have that
%\beq {\partial Z\over\partial z}=2 z {\partial Z\over \partial z^2}= 8 z\delta g^2 \mathfrak{m}^2\, %v^2\left[z^6+{v^2\over 2}
%\left(1+\delta g\right)^2z^2-\delta h{\delta g-3\over 2\delta g}\right]\,.
%\eeq{derZ}
%It is simple algebra to show that there exist no real, positive and greater than one solutions to~(\ref{derZ})

Next, we would like to show that $\left(\partial {\mathbb G}\over\partial u\right)$ and
$\left(\partial {\mathbb G}\over\partial v\right)$ are non-vanishing when evaluated at the turning point $z_t$.
Recall that we consider the turning point $z_t$ which is closest to the boundary $z=\infty$ and where
\beq p^2(z_t)\equiv \tfrac{1}{2}\left(-\left[a(z)c(z)-b(z)-d(z)\right]\pm\sqrt{\left[a(z)c(z)-b(z)-d(z)\right]^2-4b(z)d(z)}\right)_{z=z_t}=0 \,.
\eeq{p2s}
The explicit form of the functions $a(z)$, $b(z)$, $c(z)$, $d(z)$ can be found in Appendix~\ref{sec:EoM}.
Given that these functions are real, Eq.~(\ref{p2s}) essentially implies that the turning point will be a solution of the equation $b(z)d(z)=0$.

The determinant $\mathbb G$ is now defined as
\beq {\mathbb G}\equiv p^4+p^2\left[a(z)c(z)-b(z)-d(z)\right]+b(z) d(z)\,,
\eeq{Gdef2}
 It follows that
\beq  {\partial \mathbb G\over \partial u}=p^2 {\partial \over\partial u}\left[a(z)c(z)-b(z)-d(z)\right]+
{\partial \over\partial u}\, b(z)d(z).
\eeq{dGdu}
Evaluated at the turning point $z=z_t$, where $p^2(z_t)=0$, Eq.~(\ref{dGdu}) reduces to
\beq \left(\partial \mathbb G\over \partial u\right) =\left. {\partial \over\partial u}\, b(z)d(z)\right|_{z=z_t}\,.
\eeq{dGdutp}
With the help of Appendix~\ref{sec:EoM}, it is easy to deduce that
\beq b(z)d(z)=P^{-1}(z)\left[u^4 P_4(z)+u^3 P_3(z)+u^2 P_2(z)+u P_1(z)+P_0(z)\right],
\eeq{bd1}
where $P(z),\,P_i(z)$ with $i=1,...,4$ are polynomials of $z$, and $P(z)$ diverges like $z^8$ at the boundary.
Since $b(z)d(z)$ is a polynomial in $u$ it can also be written as
\beq b(z)d(z)=P^{-1}(z)P_4(z)\left(u-u_1(z)\right)\left(u-u_2(z)\right)\left(u-u_3(z)\right)\left(u-u_4(z)\right).
\eeq{db2}
At the turning point under consideration some but not all of the parentheses may vanish
so that~(\ref{dGdutp}) is finite and different from zero.
In a similar manner one can show that ${\partial {\mathbb G}\over\partial v}\neq 0$.
Finally let us note for completeness that ${\partial \mathbb G\over \partial p}$ behaves like $z^{-8}$ close to the
boundary and thus ${\partial p\over\partial u}$ is finite there.

\end{appendix}


\begin{thebibliography}{99}

%\cite{Maldacena:1997re}
\bibitem{Maldacena:1997re}
  J.~M.~Maldacena,
  %``The Large N limit of superconformal field theories and supergravity,''
  Adv.\ Theor.\ Math.\ Phys.\  {\bf 2}, 231 (1998)
  [hep-th/9711200].
  %%CITATION = HEP-TH/9711200;%%

%\cite{Gubser:1998bc}
\bibitem{Gubser:1998bc}
  S.~S.~Gubser, I.~R.~Klebanov and A.~M.~Polyakov,
  %``Gauge theory correlators from noncritical string theory,''
  Phys.\ Lett.\ B {\bf 428}, 105 (1998)
  [hep-th/9802109].
  %%CITATION = HEP-TH/9802109;%%

%\cite{Witten:1998qj}
\bibitem{Witten:1998qj}
  E.~Witten,
  %``Anti-de Sitter space and holography,''
  Adv.\ Theor.\ Math.\ Phys.\  {\bf 2}, 253 (1998)
  [hep-th/9802150].
  %%CITATION = HEP-TH/9802150;%%

\bibitem{PM1}
  S.~Deser and A.~Waldron,
  %``Null propagation of partially massless higher spins in (A)dS and   cosmological constant speculations,''
  Phys.\ Lett.\ B {\bf 513}, 137 (2001)
  [hep-th/0105181];
  %%CITATION = HEP-TH/0105181;%%
  %``Partial masslessness of higher spins in (A)dS,''
  Nucl.\ Phys.\ B {\bf 607}, 577 (2001)
  [hep-th/0103198];
  %%CITATION = HEP-TH/0103198;%%
  %``Stability of massive cosmological gravitons,''
  Phys.\ Lett.\ B {\bf 508}, 347 (2001)
  [hep-th/0103255];
  %%CITATION = HEP-TH/0103255;%%
  %``Gauge invariances and phases of massive higher spins in (A)dS,''
  Phys.\ Rev.\ Lett.\  {\bf 87}, 031601 (2001)
  [hep-th/0102166].
  %%CITATION = HEP-TH/0102166;%%
  S.~Deser and R.~I.~Nepomechie,
  %``Gauge Invariance Versus Masslessness In De Sitter Space,''
  Annals Phys.\  {\bf 154}, 396 (1984);
  %%CITATION = APNYA,154,396;%%
  %``Anomalous Propagation Of Gauge Fields In Conformally Flat Spaces,''
  Phys.\ Lett.\ B {\bf 132}, 321 (1983).
  %%CITATION = PHLTA,B132,321;%%

\bibitem{PM2}
  A.~Higuchi,
  %``Massive Symmetric Tensor Field In Space-times With A Positive  Cosmological Constant,''
  Nucl.\ Phys.\ B {\bf 325}, 745 (1989);
  %%CITATION = NUPHA,B325,745;%%
  %``Forbidden Mass Range For Spin-2 Field Theory In De Sitter Space-time,''
  Nucl.\ Phys.\ B {\bf 282}, 397 (1987);
  %%CITATION = NUPHA,B282,397;%%
%``Massive Symmetric Tensor Field In Curved Space-time,''
  Class.\ Quant.\ Grav.\  {\bf 6}, 397 (1989).
  %%CITATION = CQGRD,6,397;%%

\bibitem{PM3}
  I.~L.~Buchbinder, D.~M.~Gitman and V.~D.~Pershin,
  %``Causality of massive spin-2 field in external gravity,''
  Phys.\ Lett.\ B {\bf 492}, 161 (2000)
  [hep-th/0006144].
  %%CITATION = HEP-TH/0006144;%%
  I.~L.~Buchbinder, D.~M.~Gitman, V.~A.~Krykhtin and V.~D.~Pershin,
  %``Equations of motion for massive spin-2 field coupled to gravity,''
  Nucl.\ Phys.\ B {\bf 584}, 615 (2000)
  [hep-th/9910188].
  %%CITATION = HEP-TH/9910188;%%

\bibitem{Dolan}
  L.~Dolan, C.~R.~Nappi and E.~Witten,
  %``Conformal operators for partially massless states,''
  JHEP {\bf 0110}, 016 (2001)  [hep-th/0109096].
  %%CITATION = HEP-TH/0109096;%%

%\cite{Lovelock:1971yv}
\bibitem{Lovelock:1971yv}
   C.~Lanczos
   Z.\ Phys.\  {\bf 73}, 147 (1932).
   C.~Lanzcos
   Ann.\ Math.\  {\bf 39}, 842 (1938).
  D.~Lovelock,
  %``The Einstein tensor and its generalizations,''
  J.\ Math.\ Phys.\  {\bf 12}, 498 (1971).
  %%CITATION = JMAPA,12,498;%%


\bibitem{Myers}
  M.~Brigante, H.~Liu, R.~C.~Myers, S.~Shenker and S.~Yaida,
  %``The Viscosity Bound and Causality Violation,''
  Phys.\ Rev.\ Lett.\  {\bf 100}, 191601 (2008)  [arXiv:0802.3318 [hep-th]];
  %%CITATION = ARXIV:0802.3318;%%
  %``Viscosity Bound Violation in Higher Derivative Gravity,''
  Phys.\ Rev.\ D {\bf 77}, 126006 (2008)  [arXiv:0712.0805 [hep-th]].
  %%CITATION = ARXIV:0712.0805;%%

%\cite{Buchel:2009tt}
\bibitem{Buchel:2009tt}
  A.~Buchel and R.~C.~Myers,
  %``Causality of Holographic Hydrodynamics,''
  JHEP {\bf 0908}, 016 (2009)
  [arXiv:0906.2922 [hep-th]].
  %%CITATION = ARXIV:0906.2922;%%

%\cite{Hofman:2009ug}
\bibitem{Hofman:2009ug}
  D.~M.~Hofman,
  %``Higher Derivative Gravity, Causality and Positivity of Energy in a UV complete QFT,''
  Nucl.\ Phys.\ B {\bf 823}, 174 (2009)
  [arXiv:0907.1625 [hep-th]].
  %%CITATION = ARXIV:0907.1625;%%



\bibitem{GB}
  A.~Buchel, J.~Escobedo, R.~C.~Myers, M.~F.~Paulos, A.~Sinha and M.~Smolkin,
  %``Holographic GB gravity in arbitrary dimensions,''
  JHEP {\bf 1003}, 111 (2010)
  [arXiv:0911.4257 [hep-th]].
  %%CITATION = ARXIV:0911.4257;%%
  J.~de Boer, M.~Kulaxizi and A.~Parnachev,
  %``AdS(7)/CFT(6), Gauss-Bonnet Gravity, and Viscosity Bound,''
  JHEP {\bf 1003}, 087 (2010)  [arXiv:0910.5347 [hep-th]].
  %%CITATION = ARXIV:0910.5347;%%
  X.~O.~Camanho and J.~D.~Edelstein,
  %``Causality constraints in AdS/CFT from conformal collider physics and Gauss-Bonnet gravity,''
  JHEP {\bf 1004}, 007 (2010)
  [arXiv:0911.3160 [hep-th]].
  %%CITATION = ARXIV:0911.3160;%%


\bibitem{Lovelock}
  J.~de Boer, M.~Kulaxizi and A.~Parnachev,
  %``Holographic Lovelock Gravities and Black Holes,''
  JHEP {\bf 1006}, 008 (2010)  [arXiv:0912.1877 [hep-th]].
  %%CITATION = ARXIV:0912.1877;%%
  X.~O.~Camanho and J.~D.~Edelstein,
  %``Causality in AdS/CFT and Lovelock theory,''
  JHEP {\bf 1006}, 099 (2010)  [arXiv:0912.1944 [hep-th]].
  %%CITATION = ARXIV:0912.1944;%%
  X.~O.~Camanho, J.~D.~Edelstein and M.~F.~Paulos,
  %``Lovelock theories, holography and the fate of the viscosity bound,''
  JHEP {\bf 1105}, 127 (2011)
  [arXiv:1010.1682 [hep-th]].
  %%CITATION = ARXIV:1010.1682;%%
  R.~C.~Myers, M.~F.~Paulos and A.~Sinha,
  %``Holographic studies of quasi-topological gravity,''
  JHEP {\bf 1008}, 035 (2010)
  [arXiv:1004.2055 [hep-th]].
  %%CITATION = ARXIV:1004.2055;%%


%\cite{Hofman:2008ar}
\bibitem{Hofman:2008ar}
  D.~M.~Hofman and J.~Maldacena,
  %``Conformal collider physics: Energy and charge correlations,''
  JHEP {\bf 0805}, 012 (2008)
  [arXiv:0803.1467 [hep-th]].
  %%CITATION = ARXIV:0803.1467;%%

%\cite{Kulaxizi:2010jt}
\bibitem{Kulaxizi:2010jt}
  M.~Kulaxizi and A.~Parnachev,
  %``Energy Flux Positivity and Unitarity in CFTs,''
  Phys.\ Rev.\ Lett.\  {\bf 106}, 011601 (2011)
  [arXiv:1007.0553 [hep-th]].
  %%CITATION = ARXIV:1007.0553;%%



\bibitem{Lorce}
  C.~Lorce,
  %``Electromagnetic properties for arbitrary spin particles: Natural electromagnetic moments from light-cone arguments,''
  Phys.\ Rev.\ D {\bf 79}, 113011 (2009)  [arXiv:0901.4200 [hep-ph]].
  %%CITATION = ARXIV:0901.4200;%%

\bibitem{Aronson}
  H.~Aronson,
  %``Spin-1 electrodynamics with an electric quadrupole moment,''
  Phys.\ Rev.\  {\bf 186}, 1434 (1969).
  %%CITATION = PHRVA,186,1434;%%

\bibitem{VZ}
  G.~Velo and D.~Zwanziger,
  %``Noncausality and other defects of interaction lagrangians for particles with spin one and higher,''
  Phys.\ Rev.\  {\bf 188}, 2218 (1969).
  %%CITATION = PHRVA,188,2218;%%

\bibitem{BH}
  S.~J.~Brodsky and J.~R.~Hiller,
  %``Universal properties of the electromagnetic interactions of spin one systems,''
  Phys.\ Rev.\ D {\bf 46}, 2141 (1992).
  %%CITATION = PHRVA,D46,2141;%%

\bibitem{Tiktopoulos}
  J.~M.~Cornwall, D.~N.~Levin and G.~Tiktopoulos,
  %``Uniqueness of spontaneously broken gauge theories,''
  Phys.\ Rev.\ Lett.\  {\bf 30}, 1268 (1973)  [Erratum-ibid.\  {\bf 31}, 572 (1973)];
  %%CITATION = PRLTA,30,1268;%%
  %``Derivation of Gauge Invariance from High-Energy Unitarity Bounds on the s Matrix,''
  Phys.\ Rev.\ D {\bf 10}, 1145 (1974)  [Erratum-ibid.\ D {\bf 11}, 972 (1975)].
  %%CITATION = PHRVA,D10,1145;%%

\bibitem{SM}
  W.~A.~Bardeen, R.~Gastmans and B.~E.~Lautrup,
  %``Static quantities in Weinberg's model of weak and electromagnetic interactions,''
  Nucl.\ Phys.\ B {\bf 46}, 319 (1972).
  %%CITATION = NUPHA,B46,319;%%

\bibitem{Kim}
  K.~J.~Kim and Y.~-S.~Tsai,
  %``Magnetic Dipole And Electric Quadrupole Moments Of W+- Meson,''
  Phys.\ Rev.\ D {\bf 7}, 3710 (1973).
  %%CITATION = PHRVA,D7,3710;%%

\bibitem{Weinberg}
  S.~Weinberg,
  \emph{Lectures On Elementary Particles And Quantum Field Theory},
  Proceedings of the Summer Institute, Brandeis University, 1970, Vol. I,
  (MIT Press, Cambridge, MA, 1970).

\bibitem{g=2}
  S.~Ferrara, M.~Porrati and V.~L.~Telegdi,
  %``g = 2 as the natural value of the tree level gyromagnetic ratio of elementary particles,''
  Phys.\ Rev.\ D {\bf 46}, 3529 (1992).
  %%CITATION = PHRVA,D46,3529;%%

\bibitem{EM-SUSY}
  S.~Ferrara and M.~Porrati,
  %``Supersymmetric sum rules on magnetic dipole moments of arbitrary spin particles,''
  Phys.\ Lett.\ B {\bf 288}, 85 (1992).
  %%CITATION = PHLTA,B288,85;%%

\bibitem{Strings}
  P.~C.~Argyres and C.~R.~Nappi,
  %``Massive Spin-2 Bosonic String States In An Electromagnetic Background,''
  Phys.\ Lett.\ B {\bf 224}, 89 (1989).
  %%CITATION = PHLTA,B224,89;%%
  M.~Porrati, R.~Rahman and A.~Sagnotti,
  %``String Theory and The Velo-Zwanziger Problem,''
  Nucl.\ Phys.\ B {\bf 846}, 250 (2011)  [arXiv:1011.6411 [hep-th]].
  %%CITATION = ARXIV:1011.6411;%%

\bibitem{VZ1}
  G.~Velo and D.~Zwanziger,
  %``Propagation And Quantization Of Rarita-Schwinger Waves In An External
  %Electromagnetic Potential,''
  Phys.\ Rev.\  {\bf 186}, 1337 (1969).
  %%CITATION = PHRVA,186,1337;%%
  G.~Velo,
  %``Anomalous behaviour of a massive spin two charged particle in an external electromagnetic field,''
  Nucl.\ Phys.\ B {\bf 43}, 389 (1972).  %%CITATION = NUPHA,B43,389;%%

\bibitem{Misc}
  A.~Shamaly, A.~Z.~Capri,
  %``Propagation of interacting fields,''
  Annals Phys.\  {\bf 74}, 503-523 (1972);
  %``Electrodynamics and propagation of a spin 3/2 field,''
  Can.\ J.\ Phys.\  {\bf 52}, 919-920 (1974).
  M.~Hortacsu,
  %``Demonstration of noncausality for the rarita-schwinger equation,''
  Phys.\ Rev.\  D {\bf 9}, 928 (1974).
  %%CITATION = PHRVA,D9,928;%%
  J.~Prabhakaran, M.~Seetharaman, P.~M.~Mathews,
  %``Causality of Propagation of Spin 3/2 Fields Coupled to Spinor and Scalar Fields,''
  Phys.\ Rev.\  {\bf D12}, 3191-3194 (1975);
  %``Rarita-Schwinger Particles in Homogeneous Magnetic Fields, and Inconsistencies of Spin 3/2 Theories,''
  Phys.\ Rev.\  {\bf D12}, 458-466 (1975);
  %``Causality and Indefiniteness of Charge in Spin 3/2 Field Theories,''
  J.\ Phys.\ A {\bf A8}, 560-565 (1975).

\bibitem{DW}
  S.~Deser, V.~Pascalutsa and A.~Waldron,
  %``Massive spin 3/2 electrodynamics,''
  Phys.\ Rev.\  D {\bf 62}, 105031 (2000)
  [arXiv:hep-th/0003011].
  %%CITATION = PHRVA,D62,105031;%%
  S.~Deser and A.~Waldron,
  %``Inconsistencies of massive charged gravitating higher spins,''
  Nucl.\ Phys.\  B {\bf 631}, 369 (2002)
  [arXiv:hep-th/0112182].
  %%CITATION = NUPHA,B631,369;%%

\bibitem{PR2}
  M.~Porrati and R.~Rahman,
  %``Causal Propagation of a Charged Spin 3/2 Field in an External Electromagnetic Background,''
  Phys.\ Rev.\ D {\bf 80}, 025009 (2009)  [arXiv:0906.1432 [hep-th]].
  %%CITATION = ARXIV:0906.1432;%%
  R.~Rahman,
  %``Helicity-1/2 Mode as a Probe of Interactions of Massive Rarita-Schwinger Field,''
  arXiv:1111.3366 [hep-th].
  %%CITATION = ARXIV:1111.3366;%%

\bibitem{GR-SUSY}
  I.~Giannakis, J.~T.~Liu and M.~Porrati,
  %``Supersymmetry and gravitational quadrupoles,''
  Phys.\ Lett.\ B {\bf 469}, 129 (1999)  [hep-th/9909012].
  %%CITATION = HEP-TH/9909012;%%

\bibitem{Khriplovich}
  I.~B.~Khriplovich,
  %``Particle with internal angular momentum in a gravitational field,''
  Sov.\ Phys.\ JETP {\bf 69}, 217 (1989)  [Zh.\ Eksp.\ Teor.\ Fiz.\  {\bf 96}, 385 (1989)].
  %%CITATION = SPHJA,69,217;%%
  I.~B.~Khriplovich and A.~A.~Pomeransky,
  %``Equations of motion of spinning relativistic particle in external fields,''
  J.\ Exp.\ Theor.\ Phys.\  {\bf 86}, 839 (1998)  [Zh.\ Eksp.\ Teor.\ Fiz.\  {\bf 113}, 1537 (1998)]  [gr-qc/9710098];
  %%CITATION = GR-QC/9710098;%%
  %``Equations of motion of spinning relativistic particle in external fields,''
  Surveys High Energ.\ Phys.\  {\bf 14}, 145 (1999)  [gr-qc/9809069].
  %%CITATION = GR-QC/9809069;%%

\bibitem{h=1}
  M.~Porrati,
  %``Massive spin 5/2 fields coupled to gravity: Tree level unitarity versus the equivalence principle,''
  Phys.\ Lett.\ B {\bf 304}, 77 (1993)  [gr-qc/9301012].
  %%CITATION = GR-QC/9301012;%%
  A.~Cucchieri, M.~Porrati and S.~Deser,
  %``Tree level unitarity constraints on the gravitational couplings of higher spin massive fields,''
  Phys.\ Rev.\ D {\bf 51}, 4543 (1995)  [hep-th/9408073].
  %%CITATION = HEP-TH/9408073;%%

\bibitem{AdSRN}
  L.~J.~Romans,
  %``Supersymmetric, cold and lukewarm black holes in cosmological Einstein-Maxwell theory,''
  Nucl.\ Phys.\ B {\bf 383}, 395 (1992)  [hep-th/9203018].
  %%CITATION = HEP-TH/9203018;%%
  A.~Chamblin, R.~Emparan, C.~V.~Johnson and R.~C.~Myers,
  %``Charged AdS black holes and catastrophic holography,''
  Phys.\ Rev.\ D {\bf 60}, 064018 (1999)  [hep-th/9902170].
  %%CITATION = HEP-TH/9902170;%%

%%%%%%%%%%%%%%%%%%%%%%%%%%%%%%%%%%%%%%%%%%%%%%%%%%%%%%%%%%%%%%%%%%%%%%%%%



\bibitem{wkb}
  K.~Yabana and H.~Horiuchi,
  %``Relativistic Chiral Mean Field Model for Finite Nuclei,''
  Prog.\ Theor.\ Phys.\  {\bf 71}, 1275 (1984);
  %[arXiv:1209.0874 [nucl-th]].
  %%CITATION = ARXIV:1209.0874;%%
  %``Relativistic Chiral Mean Field Model for Finite Nuclei,''
  Prog.\ Theor.\ Phys.\  {\bf 75}, 592 (1986).
  %[arXiv:1209.0874 [nucl-th]].
  %%CITATION = ARXIV:1209.0874;%%

\bibitem{Faulkner:2009wj}
  T.~Faulkner, H.~Liu, J.~McGreevy and D.~Vegh,
  %``Emergent quantum criticality, Fermi surfaces, and AdS(2),''
  Phys.\ Rev.\ D {\bf 83}, 125002 (2011)
  [arXiv:0907.2694 [hep-th]].
  %%CITATION = ARXIV:0907.2694;%%

\bibitem{Myers:2007we}
  R.~C.~Myers, A.~O.~Starinets and R.~M.~Thomson,
  %``Holographic spectral functions and diffusion constants for fundamental matter,''
  JHEP {\bf 0711}, 091 (2007)
  [arXiv:0706.0162 [hep-th]].
  %%CITATION = ARXIV:0706.0162;%%

\bibitem{Stability}
  X.~H.~Ge and S.~J.~Sin,
  %``Shear viscosity, instability and the upper bound of the Gauss-Bonnet coupling constant,''
  JHEP {\bf 0905}, 051 (2009)
  [arXiv:0903.2527 [hep-th]].
  %%CITATION = ARXIV:0903.2527;%%
  X.~H.~Ge, Y.~Matsuo, F.~W.~Shu, S.~J.~Sin and T.~Tsukioka,
  %``Viscosity Bound, Causality Violation and Instability with Stringy Correction and Charge,''
  JHEP {\bf 0810}, 009 (2008)
  [arXiv:0808.2354 [hep-th]].
  %%CITATION = ARXIV:0808.2354;%%
  R.~G.~Cai, Z.~Y.~Nie and Y.~W.~Sun,
  %``Shear Viscosity from Effective Couplings of Gravitons,''
  Phys.\ Rev.\ D {\bf 78}, 126007 (2008)
  [arXiv:0811.1665 [hep-th]].
  %%CITATION = ARXIV:0811.1665;%%
  R.~G.~Cai, Z.~Y.~Nie, N.~Ohta and Y.~W.~Sun,
  %``Shear Viscosity from Gauss-Bonnet Gravity with a Dilaton Coupling,''
  Phys.\ Rev.\ D {\bf 79}, 066004 (2009)
  [arXiv:0901.1421 [hep-th]].
  %%CITATION = ARXIV:0901.1421;%%

%%%%%%%%%%%%%%%%%%%%%%%%%%%%%%%%%%%%%%%%%%%%%%%%%%%%%%%%%%%%%%%%%%%%%%%%%%%%%%%%%%%%%%%%%%%%%%%%%%%%%%%
%%%%%%%%%%%%%%%%%%%%%%%%%%%%%%%%%%%%%%%%%%%%%%%%%%%%%%%%%%%%%%%%%%%%%%%%%%%%%%%%%%%%%%%%%%%%%%%%%%%%%%%
%%%%%%%%%%%%%%%%%%%%%%%%%%%%%%%%%%%%%%%%%%%%%%%%%%%%%%%%%%%%%%%%%%%%%%%%%%%%%%%%%%%%%%%%%%%%%%%%%%%%%%%

\bibitem{Perlmutter}
  E.~Perlmutter, T.~Prochazka and J.~Raeymaekers,
  %``The semiclassical limit of W_N CFTs and Vasiliev theory,''
  arXiv:1210.8452 [hep-th].
  %%CITATION = ARXIV:1210.8452;%%

%\cite{Ammon:2010pg}
\bibitem{Ammon:2010pg}
  M.~Ammon, J.~Erdmenger, M.~Kaminski and A.~O'Bannon,
  %``Fermionic Operator Mixing in Holographic p-wave Superfluids,''
  JHEP {\bf 1005}, 053 (2010)
  [arXiv:1003.1134 [hep-th]].
  %%CITATION = ARXIV:1003.1134;%%


\bibitem{CT}
  I.~Bah, A.~Faraggi, J.~I.~Jottar, R.~G.~Leigh and L.~A.~Pando Zayas,
  %``Fermions and $D=11$ Supergravity On Squashed Sasaki-Einstein Manifolds,''
  JHEP {\bf 1102}, 068 (2011)
  [arXiv:1008.1423 [hep-th]].
  %%CITATION = ARXIV:1008.1423;%%
  I.~Bah, A.~Faraggi, J.~I.~Jottar and R.~G.~Leigh,
  %``Fermions and Type IIB Supergravity On Squashed Sasaki-Einstein Manifolds,''
  JHEP {\bf 1101}, 100 (2011)
  [arXiv:1009.1615 [hep-th]].
  %%CITATION = ARXIV:1009.1615;%%


\bibitem{DeWolfeab}
  O.~DeWolfe, S.~S.~Gubser and C.~Rosen,
  %``Fermi Surfaces in Maximal Gauged Supergravity,''
  Phys.\ Rev.\ Lett.\  {\bf 108}, 251601 (2012)
  [arXiv:1112.3036 [hep-th]].
  %%CITATION = ARXIV:1112.3036;%%
  %8 citations counted in INSPIRE as of 05 Mar 2013
  O.~DeWolfe, S.~S.~Gubser and C.~Rosen,
  %``Fermi surfaces in N=4 Super-Yang-Mills theory,''
  Phys.\ Rev.\ D {\bf 86}, 106002 (2012)
  [arXiv:1207.3352 [hep-th]].
  %%CITATION = ARXIV:1207.3352;%%
  %4 citations counted in INSPIRE as of 05 Mar 2013



\bibitem{MT}
   K.~-Y.~Kim and M.~Taylor,
  %``Holographic d-wave superconductors,''
  arXiv:1304.6729 [hep-th].
  %%CITATION = ARXIV:1304.6729;%%
   M.~Taylor,
   Talk at ``Holography, Gauge Theory and Black Holes" Workshop,
   Amsterdam, 17-19 Dec. 2012.


\bibitem{spin1/2}
  M.~Edalati, R.~G.~Leigh and P.~W.~Phillips,
  %``Dynamically Generated Mott Gap from Holography,''
  Phys.\ Rev.\ Lett.\  {\bf 106}, 091602 (2011)
  [arXiv:1010.3238 [hep-th]].
  %%CITATION = ARXIV:1010.3238;%%
  M.~Edalati, R.~G.~Leigh, K.~W.~Lo and P.~W.~Phillips,
  %``Dynamical Gap and Cuprate-like Physics from Holography,''
  Phys.\ Rev.\ D {\bf 83}, 046012 (2011)
  [arXiv:1012.3751 [hep-th]].
  %%CITATION = ARXIV:1012.3751;%%
  D.~Guarrera and J.~McGreevy,
  %``Holographic Fermi surfaces and bulk dipole couplings,''
  arXiv:1102.3908 [hep-th].
  %%CITATION = ARXIV:1102.3908;%%

\bibitem{d-wave}
  F.~Benini, C.~P.~Herzog, R.~Rahman and A.~Yarom,
  %``Gauge gravity duality for d-wave superconductors: prospects and challenges,''
  JHEP {\bf 1011}, 137 (2010)  [arXiv:1007.1981 [hep-th]].
  %%CITATION = ARXIV:1007.1981;%%

\bibitem{BFbound}
  P.~Breitenlohner and D.~Z.~Freedman,
  %``Positive Energy in anti-De Sitter Backgrounds and Gauged Extended Supergravity,''
  Phys.\ Lett.\ B {\bf 115}, 197 (1982).
  %%CITATION = PHLTA,B115,197;%%

\end{thebibliography}
\end{document}